# Clean few-cycle blue soliton self-compressed pulses generation in hollow-core fibers

Ziping Huang, Jiale Peng, Weitao He, Hongyu Chen, Chengbo Sun, Zhihao Wang, Shuangxi Peng, Lixin He, Qingbin Zhang*, Peixiang Lu

Wuhan National Laboratory for Optoelectronics and School of Physics, Huazhong University of Science and Technology, Wuhan 430074, PR China

**Abstract:** Blue pulses with few-cycle temporal durations hold significant value in attosecond science and ultrafast spectroscopy. In this work, we combine efficient broadband frequency doubling, multiplate continuum (MPC) post-compression and blue soliton self-compression in hollow-core fibers (HCF), experimentally demonstrating HCF-based 4.4 fs clean blue soliton self-compressed pulse. Our scheme offers three-fold advantages: (1) prevention of excessive dispersion accumulation by gradually suppressing dispersion through multi-stage design; (2) enhanced overall efficiency for self-compression-based ultrashort blue pulse generation; and (3) elimination of dispersion compensation components for the final spectral broadening stage compared to post-compression schemes. This work extends HCF-based self-compression soliton to the blue spectral region and paves the way for generating energetic ultrashort blue pulses.

## 1 Introduction

Few-cycle ultrashort pulses with exceptional peak power and intense optical fields drive critical advancements in ultrafast and strong-field research [1-3]. While commercial laser systems face intrinsic spectral bandwidth limitations, post-amplification spectral broadening techniques—collectively termed post-compression—have emerged as essential solutions [4]. At the heart of these methods lies self-phase modulation (SPM) [5], with hollow-core fiber (HCF) gas-filling nonlinear optics [6] and multiplate continuum (MPC) generation [7, 8] standing as two dominant approaches. The HCF technique, pioneered in 1996 [9], utilizes nonlinear interactions with noble gases [10] or molecular gases [11] to attain high ionization thresholds and precise dispersion control [12]. The MPC technique employs cascaded dielectric plates to deliver modular and stable spectral broadening. Both approaches, in combination with chirped-mirror dispersion compensation, are capable of generating few-cycle pulses [7, 13-15].

Alongside aforementioned chirped-mirror-assisted techniques, soliton dynamics-based pulse self-compression has emerged as an alternative approach. The pulse self-compression strategically engineers negative propagation dispersion, autonomously compensating for SPM-induced positive dispersion [16], thereby eliminates the spectral losses caused by the finite reflective bandwidth and nonlinear absorption of chirped mirrors. First proposed in micro-structured fibers [17-21], where specially designed micro-structures provide high nonlinearity and strong negative propagation dispersion to balance the SPM-induced positive dispersion. The technique nevertheless suffers from intrinsic energy scaling limitation due to its micro-structural design. To overcome this limitation, Travers et al. proposed an HCF-based 800 nm soliton self-compression scheme [16]. This scheme achieved remarkable results without external dispersion compensation, including self-compressed near-infrared sub-cycle (1.2 fs) solitons with 0.1 TW peak power, and ultraviolet (UV) resonant dispersive waves (RDW). To effectively generate the negative dispersion environments required for self-compression in HCF, pre-compressing the incident pulses (such as to 10 fs) and employing smaller inner diameters (IDs) are two typical strategies. Beyond fibers, researchers have demonstrated 4 μm self-compression in the MPC scheme [22]. However, since the negative dispersion regions of most solid-stated media are confined to the mid-infrared band, practical MPC-based self-compression at shorter wavelengths remains theoretical. Recently, Chen et al. theoretically demonstrated 800 nm self-compression in fused silica (FS) by modulating fifth-order susceptibility [23]. Consequently, HCF remains the dominant experimental platform for soliton-based pulse self-compression research.

Compared to 800 nm laser pulses, 400 nm pulses with higher single-photon energy offer unique advantages in ultrafast research. Under 400 nm pulse excitation, high-harmonic generation yields approximately 60 times more than when driven by 800 nm pulses [24]. Furthermore, 400 nm pulses compressed to 6.5 fs enable the efficient generation of isolated attosecond pulses [25, 26]. Currently, self-compression is mainly driven by infrared pulses [16, 27], the central wavelength of the RDW can only be shifted towards 400 nm by modulating the HCF's gas pressure and incident energy [28]. However, the propagation dispersion in the RDW spectral band is positive and highly nonlinear, hindering the generation of clean ultrashort temporal profiles. Additionally, the RDW spectral band lies far from the central wavelength, its generation efficiency is inherently low. Therefore, the direct efficient generation of blue self-compressed soliton with clean temporal profiles is necessary.

In this work, we present an HCF-based blue soliton self-compression scheme, achieving unprecedented performance in ultrashort pulse generation. Our approach combines efficient broadband frequency doubling of a Ti: Sapphire laser with MPC post-compression to create a 10 fs front-end source, enabling optimal conditions for blue pulse self-compression in HCF. The multi-stage architecture implements gradual dispersion compensation, effectively preventing excessive dispersion accumulation while generating clean 4.4 fs blue soliton pulses directly from the HCF output. To characterize the pulse duration, this soliton pulse is guided to a custom-built self-diffraction frequency-resolved optical gating (SD-FROG) apparatus. Since the soliton pulse propagates through additional air and window plate, the chirped mirrors are employed to compensate for the additional propagation dispersion. Although the additional propagation dispersion is well compensated, the limited bandwidth of the chirped mirrors still slightly stretches the soliton pulse to 6 fs with exceptional temporal quality (96% energy in the main pulse). Compared to infrared self-compression schemes for blue or UV RDW generation, our scheme shows higher overall efficiency while significantly reducing the energy requirements for HCF-based blue pulse generation. Notably, HCF-containing metal tube can be directly connected to subsequent chamber, external dispersion compensation in practical applications is unnecessary. Therefore, our design eliminates chirped mirrors, overcoming the reflective bandwidth constraints and nonlinear absorption effects that particularly plague blue pulse post-compression schemes. This work extends HCF-based soliton self-compression into the blue spectral region, establishing a robust platform for generating energetic ultrashort blue pulses with applications in attosecond science and precision metrology.

**2 Numerical simulations**

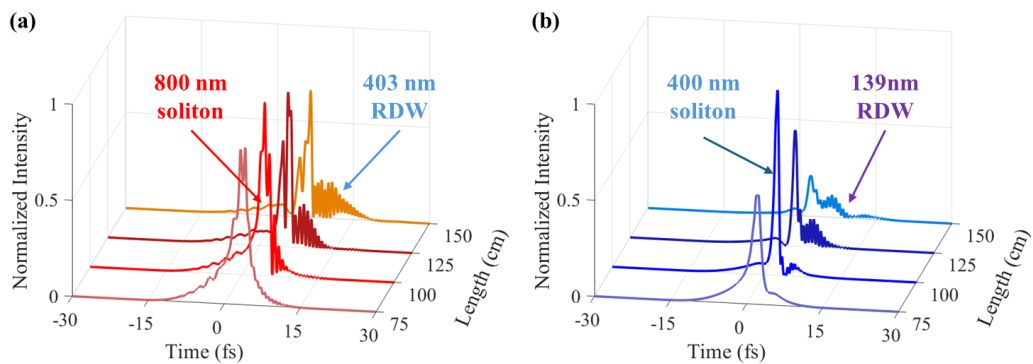

**Fig. 1.** Soliton dynamics in a gas-filled HCF. Numerically modelled generation of N = 4 solitons at 1 m in HCF filled with helium (He). (a) 470 μJ, 800 nm incident pulse generating 800 nm soliton and 403 nm RDW. (b) 26 μJ, 400 nm incident pulse generating 400 nm soliton and 139 nm RDW.

To highlight the advantages of blue soliton self-compression, we performed numerical simulations comparing infrared-soliton-based blue RDW with the direct generation of blue solitons. Soliton self-compression was simulated in a 1 m HCF by solving the unidirectional pulse propagation equation (UPPE) via a high-order Runge-Kutta method [29]. The soliton order can be modulated in the simulations by tuning the initial parameters, including the incident energy and the HCF ID [16]. Higher soliton orders result in degraded temporal profiles, whereas lower soliton orders require smaller HCF IDs, thereby increasing propagation losses. We selected $4^{th}$ order self-compressed soliton as an example. Furthermore, pre-compressing the incident pulses (such as to 10 fs) is required to establish the negative propagation dispersion environment in HCF, especially for 400 nm and 800 nm bands [16, 27]. Therefore, we first numerically modelled self-compression in HCF with 10 fs incidence, then analyzed the overall efficiency from a 30 fs Ti: Sapphire laser source to self-compression-based blue pulses. Fig. 1(a) illustrates the simulation results for 470 μJ, 10 fs pulses centered at 800 nm propagating through the HCF. After 1 m of propagation, a $4^{th}$ order 800 nm soliton and a 403 nm RDW were observed, with the blue RDW containing 10.6 μJ energy (2.2% of the 470 μJ incident energy). The front-end infrared pulse post-compression efficiency is about 60% [7], approximately 850 μJ of incident energy from the Ti: Sapphire laser source is required. Therefore, the blue RDW efficiency drops to 1.2% relative to the 850 μJ incident energy. Although extension of the propagation distance to 1.5 m (or longer) can enhance the RDW energy conversion efficiency to 2.7% (or higher), the temporal broadening and splitting occur in the RDW profile.

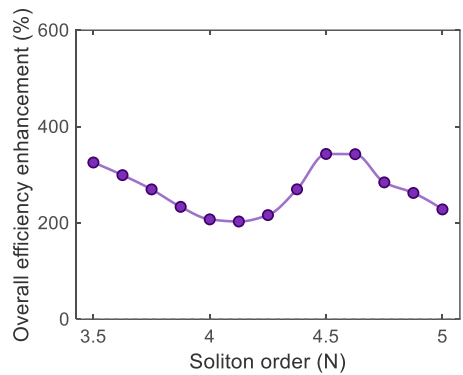

**Fig. 2.** Absolute overall efficiency enhancement ratio: 400 nm to 800 nm incidence.

In addition to the infrared pumped blue or UV RDW generation, blue soliton self-compression represents a distinct approach to generate ultrashort blue pulses. By shifting the incident central wavelength to 400 nm through front-end frequency doubling, this approach eliminates the residual energy in infrared band, which is inherent to infrared pumping. Additionally, second-harmonic generation (SHG) is a second-order nonlinear process. This gives it a fundamental advantage in energy conversion efficiency over spectrum broadening to the blue region via SPM, which is a third-order nonlinear process. As illustrated in Fig. 1(b), the propagation of 26 μJ, 10 fs pulses at 400 nm were simulated numerically in HCF, a $4^{th}$ order 400 nm soliton and a 139 nm RDW were observed at 1 m propagation distance. The blue soliton energy is 10.6 μJ (40% of the 26 μJ incidence). The front-end frequency doubling and blue pulse post-compression efficiency are about 10% and 60% respectively [30]. Thus, approximately 400 μJ of incident fundamental wave (FW) energy from the Ti: Sapphire laser source is required. The blue soliton energy conversion efficiency relative to 400 μJ incidence is 2.7%, yielding a 225% improvement over the 800 nm-pumped 1.2% blue RDW efficiency. Moreover, as illustrated in Fig.2, simulations extending to soliton orders of $3.5^{th}$-$5^{th}$ consistently confirm higher overall efficiency of the blue soliton self-compression scheme. Therefore, in comparison with the generation of blue or UV RDW via infrared pumping, our direct blue soliton scheme offers three advantages: first, it achieves at least a 2-fold enhancement in overall efficiency; second, our scheme eliminates the residual energy in the infrared band through front-end frequency doubling; third, blue soliton exhibits significantly cleaner temporal profiles than blue RDW.

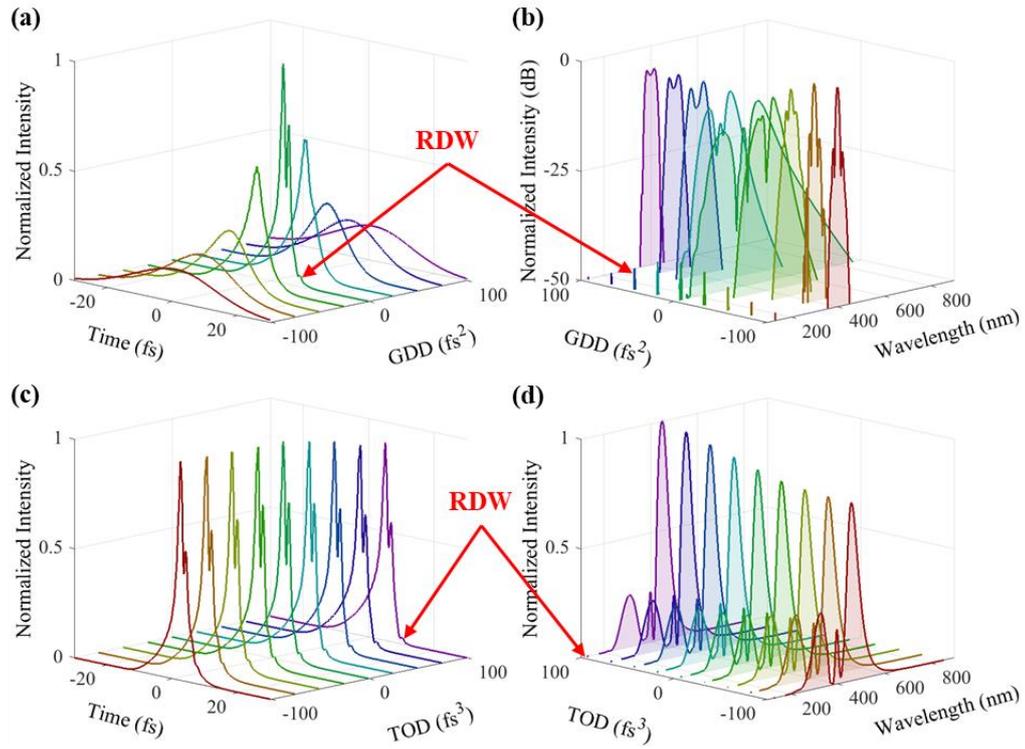

**Fig. 3.** Numerical simulated of GDD (a, b) and TOD (c, d) effects on pulse self-compression for 30 mW, 10 fs blue pulses propagation through a 100 μm ID HCF filled with 2 bar He. RDW, resonant dispersive wave.

    Residual dispersion is often present in the pre-compressed pulses experimentally, which is difficult to be fully compensated for by chirped mirrors and will propagate into the subsequent self-compression process. Therefore, we numerically analysis the blue soliton self-compression in the presence of group delay dispersion (GDD) and third-order dispersion (TOD), verifying our scheme's tolerance to dispersion and robustness under dispersion perturbation. GDD represents group velocity differences among spectral components, leading to a linear variation of frequency over time, which inevitably causes temporal broadening of the pulse and reduces the pulse peak power significantly. As illustrated in Fig. 3(a) and 3(b), simulations were performed for experimentally possible residual ±100 fs² GDD. Since SPM-induced positive dispersion is autonomously compensated in self-compression, additional pre-chirping at the front-end is unnecessary. Consequently, GDD significantly reduces the SPM strength, resulting in diminished spectral broadening, weakened RDW intensity, and symmetric broadening of the self-compressed pulse temporal profiles. While GDD can be compensated by conventional dispersion management components (e.g. chirped mirrors), managing TOD presents greater experimental challenges due to both components' limitations and higher costs. Physically, TOD introduces a linear variation in the frequency derivative, producing a parabolic frequency-to-time relationship. For instance, positive TOD retains high-frequency components at pulse edges while attenuating them at the temporal peak, enhancing long-wavelength spectral broadening. Fig. 3(c) and 3(d) confirm that ±100 fs³ TOD primarily shifts spectral broadening preferences with minimal impact on total bandwidth and RDW intensity. Crucially, TOD negligibly alters the temporal profile of blue self-compressed pulses, underscoring our scheme's robustness against higher-order dispersion.

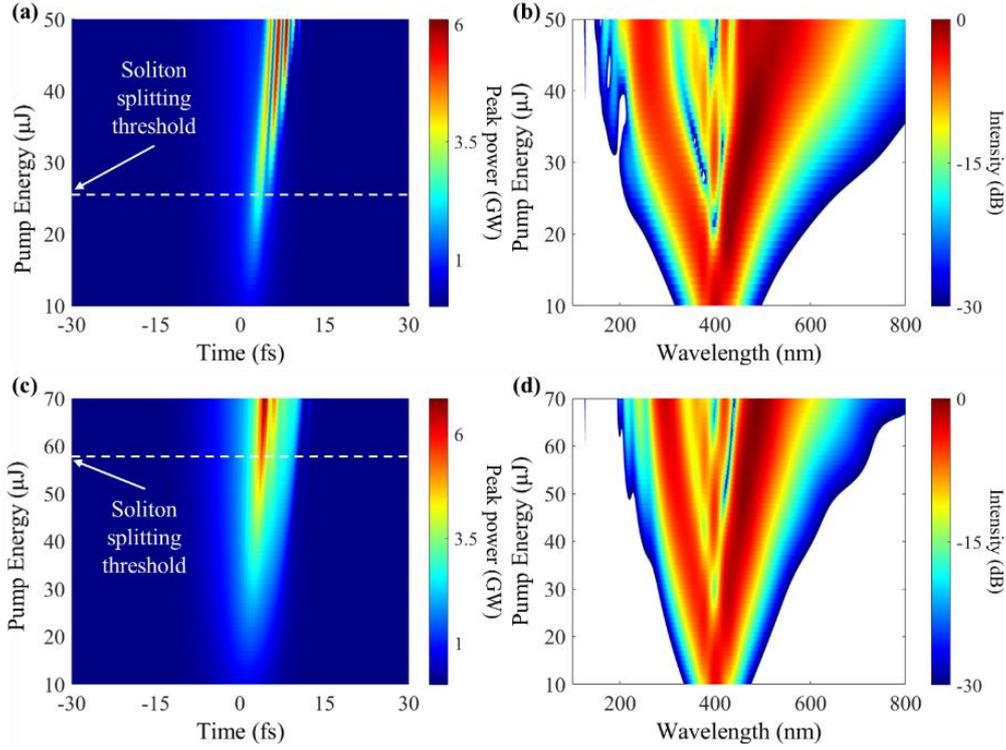

**Fig. 4.** Numerical simulated temporal (a, c) and spectral (b, d) evolutions of the 10 fs Fourier-transform-limited (FTL) blue pulse in HCF filled with 2 bar He as a function of incident energy. HCF IDs: 100 μm (a, b), 120 μm (c, d).

To explore the energy scalability of blue soliton self-compression, we numerically modelled the evolution of self-compressed pulses as the function of incident energy. Pulse self-compression arises from the balance between SPM-induced positive dispersion and the negative dispersion environment within the HCF. Notably, the latter depends on the HCF cross-sectional area, the gas type and pressure [16], remaining independent of the incident energy. Our simulations employed FTL pulse incidence while varying the incident energy, with all spectra normalized to their respective maxima. Numerical simulations of 10 fs blue pulse propagating through 2 bar He-filled HCF (100 μm ID) with different incident energies are illustrated in Fig. 4(a) and 4(b). It is evident that reduced incident energy weakens the SPM effects, leading to diminished spectral broadening. Narrower spectrum corresponds to a broader FTL pulse, resulting in a longer duration of self-compressed pulses. Conversely, elevated incident energy intensifies SPM, thereby enhancing spectral broadening and RDW energy proportion. Furthermore, enhanced SPM induces higher-order chirp. Based on the soliton dynamics principle of pulse self-compression [12], the soliton order $N = \sqrt{L_{disp}/L_{nl}}$, where $L_{disp} = (d\tau)^2/(1.763|\beta_2|)$ represents the energy-independent dispersion length, $d\tau$ is the full-width at half-maximum duration of the incident pulse, $\beta_2$ denotes the group velocity dispersion, $L_{nl} = 1/(\gamma P_0)$ is the nonlinear length, $\gamma$ is the nonlinear coefficient, and $P_0 = E_0/d\tau$ defines the peak power. From the derived relationship $N \propto \sqrt{E_0}$, excessive incident energy generates higher-order self-compressed solitons with degraded temporal profiles. In addition, the higher-order chirp induced by enhanced SPM fails to be adequately compensated for by the negative dispersion environment in HCF. This leads to pulse oscillation, triggering modulation instabilities, multiple peaks, and potential breakup into sub-pulses [31]. These detrimental effects can be weakened by implementing larger IDs. Numerical simulations of 10 fs blue pulse propagating through HCF with slightly larger ID (120 μm) are illustrated in Fig. 4(c) and 4(d). Remarkably, slightly increasing the HCF ID considerably raises the soliton splitting threshold from approximately 25 μJ to about 60 μJ. This result provides insights into the energy scalability of blue soliton self-compression scheme and suggests the feasibility of generating energetic ultrashort blue solitons.

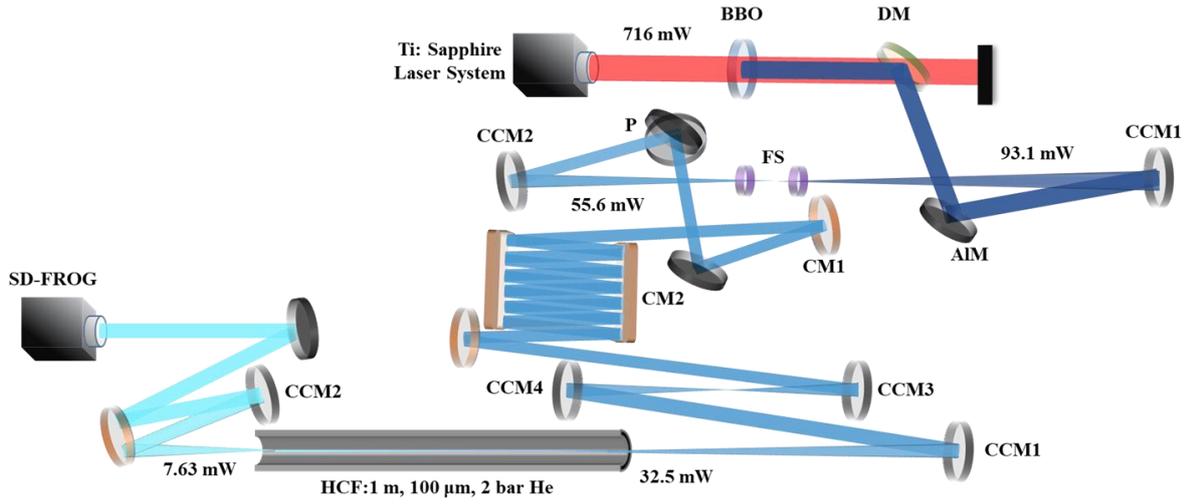

**Fig. 5.** Experimental setup. AlM, UV-enhanced Aluminum mirrors; CCM1-CCM4, UV-enhanced Aluminum-coated concave mirrors (f = 750, 500, 200, 250 mm, respectively); DM, dichroic mirror; FS, 100 μm thick fused-silica plates; P, periscope composed of two AlMs; CM1 and CM2, chirped mirrors.

## 3 Experimental results

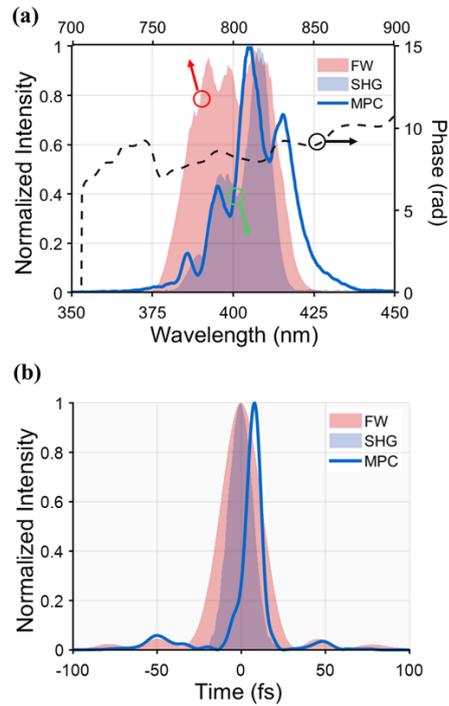

**Fig. 6.** FROG characterization of pre-compression source for self-compression. (a) Retrieved FW (red shadow), SHG (blue shadow), MPC post-compression spectrum (blue line) and phase (dashed line). (b) Retrieved FW (red shadow), SHG (blue shadow) and MPC post-compression (blue line) pulses.

The experimental setup is schematically illustrated in Fig. 5. At the optical front end, a Ti: Sapphire laser system delivering 716 mW, 29 fs infrared pulses at 1 kHz repetition rate were characterized using a custom-built second-harmonic generation FROG (SHG-FROG) apparatus. Optimization of the grating angle within the Ti: Sapphire laser system allowed tuning of the FW [7, 8]. The SHG-FROG trace retrieved in Fig. 6 (red shadow) confirms that the FW exhibited negligible high-order dispersion and near-FTL pulse duration, which is advantageous for subsequent frequency doubling and pulse compression. Frequency doubling of the FW pulse was implemented using a 72 μm thick substrate-free BBO cutting at 29.2°. As illustrated in Fig. 6 (blue shadow), we characterized the 93 mW SHG pulse using a SD-FROG apparatus, obtaining 17 fs duration (16 fs FTL pulse duration). Optimization of the laser source enabled frequency doubling with high efficiency (13%) and duration shorter than the FW pulse. The FW and SHG pulses were spatially separated using a dichroic mirror at 56° incident angle, after which the SHG pulse was directed into the MPC post-compression stage. After the SHG pulse was focused by a CCM1, two 100 μm thick FS plates were positioned before and after the focus, respectively. Although beam quality degradation occurs more easily in 400 nm MPC configurations than in those driven by longer wavelengths [32], the necessity for achieving near-10 fs duration requires placing the FS plates near the focus to ensure sufficient energy density. To balance beam quality maintenance and effective nonlinear spectral broadening, the separation between the FS plates was set to 6 cm. Following collimation via a CCM2, dispersion compensation was implemented using chirped mirrors to achieve temporal compression. The pulse underwent dispersion compensation through 1 reflection on a CM1 (Ultrafast Innovations, CM82, -50 fs$^2$ per bounce across 350-450 nm, single-bounce compatible) and 12 reflections on two CM2s (Layertec, 148920, -25 fs$^2$ per bounce across 340-400 nm), resulting in a total GDD of -350 fs$^2$. Subsequent SD-FROG characterization demonstrated a 9.96 fs post-compression blue pulse from the MPC stage, as illustrated in Fig. 6 (blue line and dashed line). The MPC stage yielded 56 mW output at 60% throughput efficiency. The energy decrease is mainly attributed to the nonlinear absorption during the 13 reflections on CM1 and CM2s.

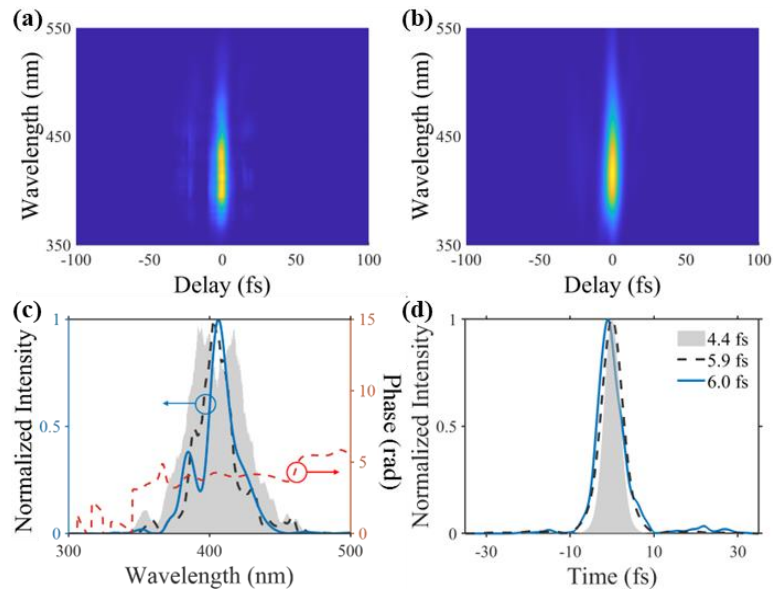

**Fig. 7.** Characterization of self-compressed pulses. Measured (a) and retrieved (b) SD-FROG trace. (c) Retrieved spectrum (blue), measured spectrum (black dashed curve), retrieved phase (red dashed curve) of self-compressed pulse, and measured spectrum of HCF direct output (shadow). (d) Retrieved temporal profile (blue), FTL pulses of self-compressed spectrum (black dashed curve) and HCF direct output (shadow).

The soliton-driven pulse self-compression process imposes critically stringent requirements on initial conditions, particularly incident pulse duration [16]. To maintain near-10 fs pulse duration at the HCF entrance, we augmented the MPC dispersion compensation scheme with an extra CM1 to deliver -50 fs$^2$ pre-chirp. This compensated for the 59 fs$^2$ GDD induced by the 0.5 mm Brewster-angled FS window placed before the HCF. Mode matching into the HCF was implemented through sequential deployment of CCM3 and CCM4 [33], followed by a loosely focused CCM1 configuration to optimize coupling efficiency for the HCF [6]. The HCF was pressurized with He at 2 bar to establish a negative dispersion environments compensating for the SPM-induced positive dispersion. The 32.5 mW, 9.96 fs, 400 nm post-compression pulse was then coupled into the HCF to initiate soliton dynamics. Direct SD-FROG characterization was performed at the HCF exit. Reconstruction analysis revealed GDD of 104.6 $fs^2$ and TOD of 28.3 $fs^3$, which were induced by propagation in air and FS window. The measured spectral range spanned 325-515 nm, supporting an FTL pulse duration of 4.4 fs. Notably, dispersion accumulation during meter-scale atmospheric propagation of few-cycle blue pulses exceeds that induced by millimeter-scale FS window transmission [20, 21, 34]. Consequently, we minimized the post-HCF optical path by two reflections (before and after collimation) on a single CM1, which compensated for the propagation dispersion. Subsequently, as illustrated in Fig. 7(d), SD-FROG characterization revealed clean blue pulses with 6 fs duration (5.9 fs FTL duration), containing 96% energy in the main pulse. Residual dispersion retrieved from the SD-FROG trace were 4.4 $fs^2$ and -13.0 $fs^3$. We then calculated the propagation dispersion from the fiber exit to the FROG's FS plate which generates the SD signal, as quantified in Table 1. After compensated for by two reflections on CM1, the residual propagation GDD and TOD were 2.6 $fs^2$ and 31.5 $fs^3$ respectively, which showed excellent agreement with the residual dispersion retrieved from SD-FROG trace. This agreement indicates that the CM1 at HCF output optical path only compensated for the propagation dispersion, proving the direct generation of clean blue soliton self-compressed pulses at the exit of HCF.

Table 1. Dispersion of medium at post-HCF optical path.

| Dispersion source | | GDD ($fs^2$) | TOD ($fs^3$) |
|---|---|---|---|
| 260 mm He | | 0.5 | 0.3 |
| 0.5 mm FS window | | 48.8 | 15.2 |
| 1065 mm air | | 53.2 | 16.0 |
| CM1 Compensation | | -100 | 0 |
| Residual dispersion | Optical path | 2.6 | 31.5 |
| | SD-FROG | 4.4 | -13.0 |

As the reflective bandwidth of the optics after HCF is narrower than the direct output spectrum, the self-compressed soliton spectrum is therefore reduced to 340-475 nm, as illustrated in Fig. 7(c). Consequently, the pulse duration characterized by SD-FROG is slightly broader than the FTL duration of HCF direct output (4.4 fs). As illustrated in Fig. 8, the waveguide and transverse mode mixing characteristics of the HCF [28, 35] yield a final output beam with 95% ellipticity and enhanced pointing stability, surpassing other pulse compression schemes such as MPC. Notably, the beam pointing exhibits periodic oscillation over time, which correlates with the temperature fluctuations in our laboratory. Directly connecting the HCF-containing metal tube to a subsequent chamber without window components enables the establishment of an lower dispersion environment [16]. This configuration eliminates the need for chirped mirrors, removing their reflective bandwidth and nonlinear absorption constraints of self-compression. Additionally, such a design provides protection for optical components against localized high energy density [7], which is particularly critical for energetic ultrashort blue pulse applications. Furthermore, implementation of the aforementioned chamber-assisted configuration in the front-end MPC pre-compression stage protects the optical components and improves beam quality. This enhancement leads to increased coupling efficiency and benefits for self-compression in HCF [35, 36].

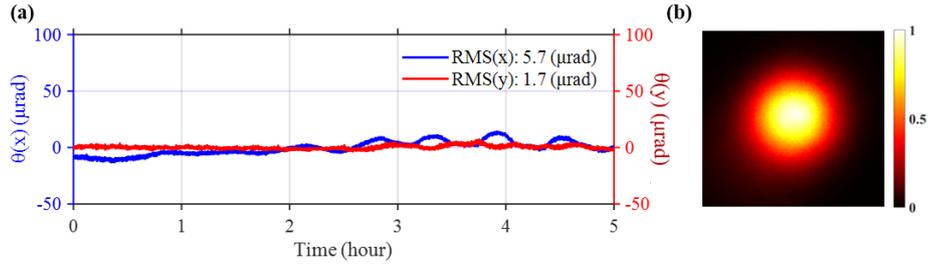

**Fig. 8.** (a) Beam-pointing stabilization of final output pulses, horizontal (blue) and vertical (red) direction. (b) Measured near-field spot of the final output beam.

**4 Conclusion**

In summary, we propose an HCF-based blue soliton self-compression pulse generation scheme. Implementing efficient broadband frequency doubling of Ti: Sapphire laser source, we achieved an SHG pulse with 13% conversion efficiency and an FTL duration shorter than the FW. Using an MPC post-compression configuration, we established a 9.96 fs blue pre-compression source for the HCF-based self-compression stage. This front-end pre-compression configuration reduces the incident energy requirement for soliton self-compression, facilitates the generation of cleaner self-compressed pulses over shorter distances, and gradually compensates dispersion to prevent its excessive accumulation. We demonstrated the generation of 4.4 fs clean blue soliton self-compressed pulses in HCF. As the characterization of the HCF output pulses was performed in atmosphere, one chirped mirror was employed to compensate for propagation dispersion induced by air and FS window. SD-FROG characterization demonstrated a near-FTL 6 fs clean blue pulse, with 96% energy contained in the main pulse. Compared to blue or UV RDW generation via infrared self-compression, our scheme enhances the overall efficiency for self-compression-based ultrashort blue pulse generation. Furthermore, our scheme theoretically eliminates the need for dispersion compensation components and their detrimental reflective bandwidth constraints and nonlinear absorption effects. These effects are especially severe in the blue spectral region and inherent in conventional post-compression schemes. Moreover, the HCF configuration ensures excellent beam ellipticity (95%) and enhanced pointing stability of the final self-compression beam. We believe this energy-efficient HCF-based blue soliton self-compression system establishes a robust platform for generating energetic ultrashort blue pulses, and holds significant potential for applications in ultrafast science, high-repetition-rate laser systems, and UV-pumped nonlinear optics research.

**Acknowledgements**

Thank the HPC Platform of Huazhong University of Science and Technology for the computation.

**Funding**

This work was supported by the Innovation Program for Quantum Science and Technology 2024ZD0300700, National Natural Science Foundation of China (NSFC) (Nos. U24A20310, 12021004), Cross Research Support Program of Huazhong University of Science and Technology (No.2023JCYJ041), Wuhan Science and Technology Major Project (No. 2024010702020023), Wuhan Semiconductor Laser Equipment Industry Innovation Joint Laboratory Project (2024050902040447), and Major Program (JD) of Hubei Province (No.203BAA015).

**Declaration of competing interest**

The authors declare that they have no known competing financial interests or personal relationships that could have appeared to influence the work reported in this paper.

**Data availability**

Data will be made available on request.